\documentclass{elsart}

\usepackage{amsmath}
\usepackage{amssymb}
\usepackage{graphicx}
\usepackage{dcolumn}
\usepackage{bm}

\def\be{\begin{eqnarray}}
\def\ee{\end{eqnarray}}
\def\ba{\begin{array}}
\def\ea{\end{array}}

\begin{document}

\begin{frontmatter}

\title{Local current distribution at large quantum dots (QDs): a
self-consistent screening model}

\author[l1]{P. M. Krishna}
\ead{phani@fen.bilkent.edu.tr},
\author[l2]{ A. Siddiki},
\author[l1]{ K. G\"uven} and
\author[l1,3]{T. Hakio\u{g}lu }

\address[l1]{Bilkent University, Department of Physics, Ankara, 06800 Turkey}
\address[l2]{Physics Department, Arnold Sommerfeld Center for Theoretical Physics, and Center for NanoScience
Ludwig-Maximilians-Universit\"at M\"unchen, D-80333 Munich,
Germany}
\address[3]{UNAM Material Science and Nanotechnology Research
Institute, Bilkent University, Ankara, 06800 Turkey}

\begin{abstract}
We report the implementation of the self-consistent Thomas-Fermi
screening theory, together with the local Ohm's law to a quantum
dot system in order to obtain local current distribution within
the dot and at the leads. We consider a large dot (size
 $>700$ nm) defined by split gates, and coupled to the leads.
Numerical calculations show that the non-dissipative current is
confined to the incompressible strips. Due to the non-linear
screening properties of the 2DES at low temperatures, this
distribution is highly sensitive to external magnetic field. Our
findings support the phenomenological models provided by the
experimental studies so far, where the formation of the (direct)
edge channels dominate the transport.
\end{abstract}
\begin{keyword}
Edge states \sep Quantum Hall effect \sep Screening \sep quantum
dots
\PACS 73.20.Dx, 73.40.Hm, 73.50.-h, 73.61.-r
\end{keyword}
\end{frontmatter}
%

\section{Introduction}
For the last two decades, the electron transport through quantum
dots (QDs) has been a central question. A wide range of different
approaches, including independent electron picture, constant
interaction model and scattering matrix theory, have provided
mechanisms for explaining the transport properties of QDs. The QDs
are constructed within a two-dimensional electron system (2DES),
by (split-) gates and/or chemical etching. In the presence of a
strong perpendicular magnetic field, it has been shown that
semi-metallic (compressible) and semi-insulating (incompressible)
regions are formed due to Coulomb
interaction~\cite{McEuen91:1926}. Despite some limited quantum
mechanical treatments, a microscopic model describing the current
distribution  is still not available for large QDs
($d>700$nm)~\cite{Keller01:qd}. In this work we implement the self
consistent (SC) Thomas-Fermi (TF) theory of
screening~\cite{Siddiki03:125315,Guven03:115327,siddiki2004}
together with the local version of the Ohm's law
~\cite{Guven03:115327,siddiki2004,TobiasK06:h} to obtain the local
current distribution inside the QD and the leads.  We use a
conductivity model~\cite{Guven03:115327,engin07} based on Gaussian
broadened density of states (following Ref.~\cite{Gross98:60}).
Our model calculations show that the non-dissipative current is,
in fact, confined to the incompressible strips. Due to the
non-linear screening properties of the 2DES at low temperatures,
this distribution is highly sensitive to the external magnetic
field and our findings support the phenomenological models
provided by the experimental groups~\cite{Keller01:qd}.
\section{Model}
Our calculation procedure starts with generating the (external)
potential $V_{\rm ext}({\bf r})$ landscape from the metallic
surface gates, which are kept at the potential $V_{\rm g}$, where
${\bf r}=(x,y)$. The calculation of $V_{\rm ext}({\bf r})$  is
based on the solution of the 2D Laplace's
equation~\cite{engin07,SiddikiMarquardt,Igor07:qpc1}. The screened
potential at zero field and zero temperature is obtained by, \be
V_{\rm scr}({\bf r})=F^{-1}[F[V_{\rm ext}({\bf r})]/\epsilon(q)],
\label{eq:vscr}\ee where $F$ presents the Fourier transformation,
 $\epsilon(q)$ is the TF dielectric function, $\epsilon(q)=1+2(m
e^2)/(\bar{\kappa} \hbar^2|q|)$, and ${\bar{\kappa}}$ is the
static dielectric constant ($\sim 12.4$ for GaAs). Including a
magnetic ($B$) field perpendicular to the 2DES, one has to solve
the Poisson's equation for the given initial potential and
electron distribution in a SC way, that is
 \be n_{\rm el}({\bf{r}})=\int dE\,D(E)f\big(
[E+V({\bf{r}})-\mu^{\star}({\bf{r}})]/k_{B}T \big),
\label{eq:tfed}\ee and \be \label{eq:hartree} V({\bf{r}})=V_{\rm
ext}({\bf{r}}) + \frac{2e^2}{\bar{\kappa}} \int_{A} \!d{\bf{r'}}
K({\bf{r}},{\bf{r'}})\,n_{\rm el}({\bf{r'}}). \ee Here, $f(\xi)$
is the Fermi function, $\mu^\star({\bf{r}})$ position dependent
chemical potential, $T$ temperature, $D(E)$ is the Gaussian
broadened Landau density of states (DOS), $V({\bf r})$ the total
potential energy, $K({\bf{r}},{\bf{r'}})$ is the Poisson kernel
satisfying periodic boundary conditions. For
 zero external current, $\mu^{\star}({\bf{r}})$ is a constant at
the thermal equilibrium, otherwise modified by the driving
electric field as \be E({\bf{r}})=\nabla
\mu^{\star}({\bf{r}})/e=\hat{\rho}({\bf{r}}).j({\bf{r}}), \ee for
a given resistivity tensor $\hat{\rho}({\bf{r}})$ and current
density $j({\bf{r}})$. We calculate the conductivity using the
Gaussian broadened DOS given by, \be D(E)=\frac{1}{2\pi
l^2}\sum_{n=0}^{\infty}
\frac{\exp(-[E_n-E]^2/\Gamma^2)}{\sqrt{\pi}\,\Gamma} \ee where
$\Gamma$ is the impurity parameter yielding the Landau level (LL)
broadening. The Landau energy is given by
$E_n=\hbar\omega_c(n+1/2)=E_{F}\Omega(n+1/2)$ where $E_F$ is the
Fermi energy. Beyond the linear response (i.e. when the current is
large enough to modify $\mu^{\star}({\bf r})$), one has to insert
the modified chemical potential into Eqn.(\ref{eq:tfed}) and
repeat the SC calculation until convergence is achieved.
\section{Results and Discussion}
\begin{figure}
{\centering
\includegraphics[width=1.0\linewidth]{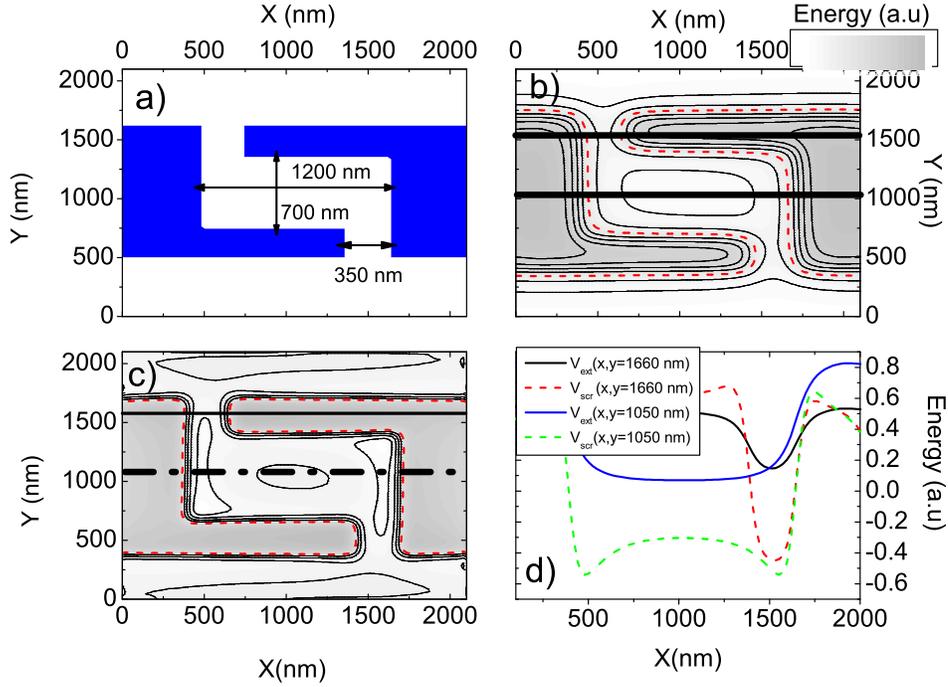}
%
\caption{ \label{fig:fig1}(a) The split gate defined Qdot.
Equipotential lines of the bare confinement (b) and screened
potential (c). Cross section of the potential profiles at the
opening and at the center of the dot (d).}}
\end{figure}
\begin{figure}
{\centering
\includegraphics[width=1.0\linewidth]{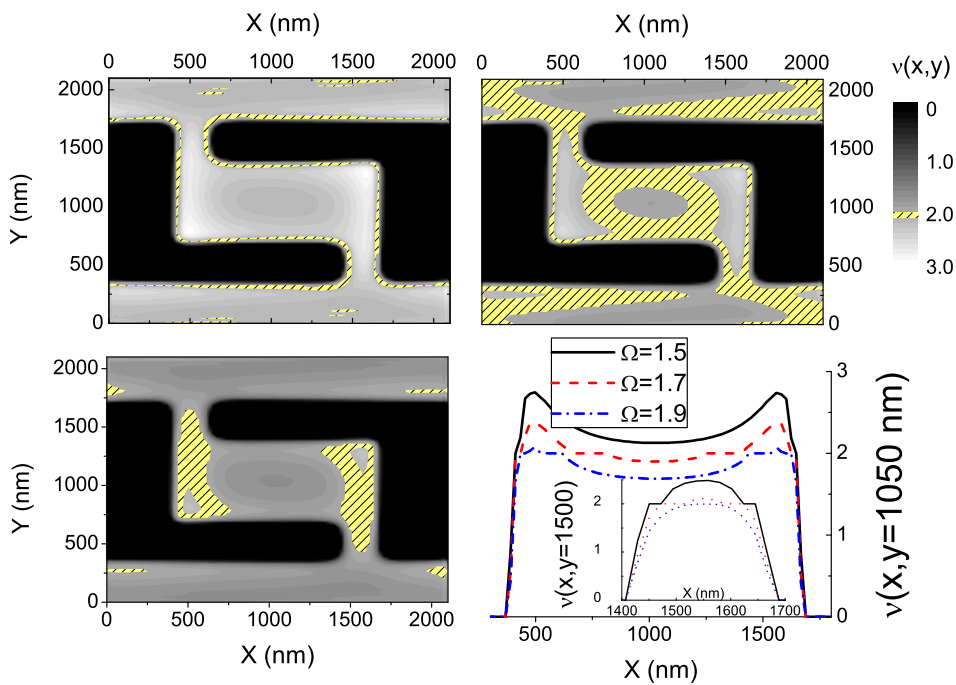}
%
\caption{ \label{fig:fig2} (a-c) The gray scale plot of $\nu(x,y)$
for three $B$ values calculated at $kT/\hbar\omega_{c}=1/40$. (d)
Density cross section again at the center (plot) and at the
opening (inset).}}
\end{figure}

We define the QD by split gates (dark (blue) regions in
Fig.\ref{fig:fig1}a) on the surface of the GaAs/AlGaAs
heterostructure, $85$ nm above the 2DES and considering a unit
cell of size $2100\times 2100$ nm$^2$, with an average Fermi
energy $12.75$ meV, corresponding a bulk electron density $3\times
10^{11}$ cm$^{-2}$ (indicated by the thick-dashed (red) contour
line). The metallic gates are biased (negatively) such that no
electrons can reside below. The QD has a rectangular shape ($\sim
1200\times 600$ nm$^2$), whereas the openings are approximately
$300$ nm. Furthermore, the QD is coupled electrically to the
leads, where the capacitive and tunnelling effects are negligible.
In Fig.\ref{fig:fig1}, we plot the bare confinement (b) and
screened (c) potentials obtained from Eqn.(\ref{eq:vscr}).
Following the contour lines, one observes that $V_{\rm ext}({\bf
r})$ is smooth, i.e. there are no potential variations within and
near the QD, whereas $V_{\rm scr}({\bf r})$ exhibits a local
extremum, due to the strong non-linear screening (for a recent
review see Ref. \cite{SiddikiMarquardt}). Two interesting
potential cross sections are highlighted in Fig.\ref{fig:fig1}d,
indicated by horizontal lines (solid, depicting the bare and
dashed the screened potential at the opening and center of the QD)
shown in the contour plots. We observe that, the screened
potential is suppressed compared to the bare potential at the
leads of the QD, implying that more states are allowed to pass
through the barrier. More interestingly, depending on the dot
size, a local maximum develops at the very center of the QD
surrounded by a minimum close to the edges where the high $q$
components dominate the screening. It is clear that, if the dot is
smaller such potential inhomogeneities will disappear, similar to
what happens at the opening, however the dot will become more
confining compared to the non-interacting models. We should also
note that, if the 2DES is buried deeper, due to exponential decay
of the short range oscillations, such local minima will not be
seen even considering smaller dot sizes. We continue our
discussion of screening now also considering an external $B$
field. In Fig.\ref{fig:fig2}, we show a sequence of local filling
factors while changing the B field. At the highest $B$ only the
lowest LL is occupied therefore the system is compressible almost
everywhere except near the gates, where local minimum is observed
(see Fig.\ref{fig:fig2}d).
In this situation, we see that the current is distributed all over
the sample, similar to a metal. From the edge state picture point
of view, a direct channel already exists and conduction is
quantized. Lowering the magnetic field results in the formation of
incompressible regions where $E_F$ falls in between two LLs. We
observe that, an incompressible ring is formed within the dot
which is connected to the leads again by incompressible strips.
Since, the current flows within the incompressible edge states,
this is the most interesting case, due to the opening of a direct
channel, which is coupled to a compressible lake inside the dot,
separated by an incompressible region. At the lowest $B$ shown in
figure, we see that the center of the dot becomes compressible
surrounded by incompressible regions, due to the local minimum
near the gates. The leads remain compressible all over, therefore
act as a metal and current is directly proportional to the local
electron density. To conclude, we have provided an explicit
calculation of the spatial distribution of the incompressible edge
states considering a split gate defined (large) QD. We have shown
that depending on the sample geometry and $B$ field applied, a
direct channel can emerge, connecting the source to the drain.
More interestingly, a dot-in-dot structure is obtained for a
certain range of parameters.

The authors acknowledge the support of the Marmaris Institute of
Theoretical and Applied Physics (ITAP), TUBITAK grant 105T110,
SFB631 and DIP.

\end{document}